\documentstyle[12pt,axodraw,a4wide]{article}
\def\eq#1{{(\ref{#1})}}

\def\tr{\mathop{\mbox{tr}}\,}

\newcommand{\eqn}[1]{(\ref{#1})}
\def\bea{\begin{eqnarray}}
\def\eea{\end{eqnarray}}
\def\be{\begin{equation}}
\def\ee{\end{equation}}

\def\r{\rho}
\def\s{\sigma}

\def\nn{\nonumber}
\newcommand{\ft}[2]{{\textstyle\frac{#1}{#2}}\,}
\newcommand{\vac}[1]{\langle #1 \rangle}

\def\v#1{{\bf v}_{#1}}
\def\q#1{{\bf q}_{#1}}
\def\r#1{{\bf r}_{#1}}
\def\b#1{{\bf b}_{#1}}
\def\OaY{Okawa and Yoneya }
\def\OY{Okawa--Yoneya }
\renewcommand{\o}{\sigma}
\begin{document}

\thispagestyle{empty}
\begin{flushright}
{\small hep-th/9905183}\\
{\small AEI-112\\
UvA-WINS-Wisk-99-09\\
NIKHEF 99-014}\\[3mm]
\end{flushright}

\vspace{1cm}
\setcounter{footnote}{0}
\begin{center}
{\Large{\bf  Three Graviton Scattering in M-Theory}
    }\\[14mm]

{\sc  Robert Helling${}^a$,
Jan Plefka${}^a$, Marco Serone${}^b$ \\and Andrew Waldron${}^c$}\\[10mm]

${}^a${\em Albert-Einstein-Institut}\\
{\em Max-Planck-Institut f\"ur
Gravitationsphysik}\\
{\em Am M\"uhlenberg 5, 14476 Potsdam-Golm, Germany}\\
{\footnotesize \tt helling@x4u2.desy.de, plefka@aei-potsdam.mpg.de}\\[7mm]

${}^b$
{\em Department of Mathematics, University of Amsterdam }\\
{\em Plantage Muidergracht 24, 1018 TV Amsterdam, The Netherlands}\\
{\footnotesize\tt serone@wins.uva.nl}\\ [7mm]

${}^c${\em NIKHEF, P.O. Box 41882, 1009 DB Amsterdam, The Netherlands}\\
{\footnotesize \tt waldron@nikhef.nl}\\[15mm]

{\sc Abstract}\\
\end{center}

The leading eikonal $S$-matrix for three graviton scattering
in $d=11$ supergravity and Matrix Theory are shown to precisely agree. The
result unifies the source-probe plus recoil approach of Okawa and
Yoneya and relaxes the restriction imposed by those authors that all
$D$-particle impact parameters and velocities are mutually
perpendicular.
Furthermore, the unified $S$-matrix approach facilitates a clean-cut
study of $M$-theoretic ${\cal R}^4$ curvature corrections to the low 
energy supergravity
effective action. In particular, the leading ${\cal R}^4$ correction
to the three graviton $S$-matrix is computed and 
compared to the corresponding next to leading order two loop $U(3)$
amplitude in Matrix Theory. We find a clear disagreement
of the two resulting tensor structures.

\vfill
\leftline{{\sc May 1999}}

\newpage
\setcounter{page}{1}

\section{Introduction}

According to current thinking, the various known string theories 
should be regarded as appropriate limits of a more fundamental
eleven dimensional theory, referred to as M-theory~\cite{witten}. 
The cornerstone of our present understanding of
M-theory is that its low energy effective action ought be
$d=11$ supergravity~\cite{Cremmer}. It has been proposed, however,  
that the quantum degrees of freedom of light-cone M-theory are
captured by a supersymmetric quantum mechanical $U(N)$ Yang--Mills model, 
known as Matrix Theory~\cite{bfss,suss}. 
In practical terms this has meant that a large
body of research has been devoted to 
comparing quantities computed via
Matrix Theory with those in $d=11$ supergravity. In particular, at the level
of comparing phase shifts for eikonal scattering of
gravitons~\cite{dkps,bbpt,oy} along with the complete tree level
$t$-channel ($2\rightarrow2$)
graviton and antisymmetric tensor $S$-matrices~\cite{psw,psw1},
impressive agreement has been found. It has also been possible to
successfully compare the conserved currents of the two models~\cite{KT}.

Nevertheless, it should be noted that computations to date have only
managed to show the equivalence of one and two loop computations 
in a relatively simple quantum mechanical model with what amounts 
to tree level supergravity. Therefore,
the capability of the Matrix Theory to uncover genuinely new physics
seems somewhat limited. This is the main question we shall address in this
paper, that is whether the model serves as a tool to study quantum
corrections to the supergravity action. To this end it is clearly
of central importance to determine the exact nature of the 
proposed correspondence. 

The first issue is to identify the correspondence between the
states of the two theories. 
Indeed, one of the motivations for the original conjecture \cite{bfss}
was the realization that the one-dimensional super Yang-Mills 
model possesses asymptotic excitations that behave as supergravitons
of eleven dimensional supergravity. This correspondence was 
refined in \cite{pw} where explicit asymptotic wave functions 
of gravitons, antisymmetric tensors and gravitini 
were found in the quantum mechanical model.
Following the lines of \cite{pw}, it has been possible 
to find a formalism to compute eikonal scattering amplitudes
for these excitations in Matrix Theory~\cite{psw,psw1}.
In this article we apply this method (which we often refer to as
the Matrix Theory LSZ formalism) to multi-particle scattering. 
In particular, we consider
three graviton amplitudes (studied already extensively within an
eikonal phase shift framework by Okawa and Yoneya~\cite{oy,oy1}).

The motivations for this computation are twofold.
Firstly, 
given the agreement found in~\cite{oy} for this process,
such a calculation
provides both a detailed test of our approach and at the same time
verifies their work. Actually our formalism will provide
not only a check of the results of~\cite{oy,oy1}, but also
an extension and unification of them.
In particular, we have been able to drop the restriction made by~\cite{oy}
that all D-particle velocities and impact parameters are 
mutually perpendicular.
Furthermore, in a direct comparison of scattering amplitudes, 
there is no need to
distinguish between recoil and non-recoil terms,
as long as one sums over all the Feynman diagrams in the theory,
including, in particular, the one particle reducible graphs.
Our result constitutes the complete agreement of $t$-channel three particle
spin independent S-matrices in  Matrix Theory and tree level
supergravity.

The second motivation of our computation arises 
from the observation of \cite{2gravR4}
that the next to leading term in the two loop effective action of Matrix
Theory, which is of order $v^8/r^{18}$ (in relative velocities and distances 
between the supergravitons), has the correct scaling to match the 
first correction to graviton scattering induced by higher order ${\cal R}^4$
curvature  
corrections~\cite{gv97} to $d=11$ supergravity.
Although this observation originally concerned two graviton scattering,
we stress that two loops in the Matrix Theory corresponds generically to three
particle interactions. Two particle scattering arises then
only as a sub-case in which the momenta of two of the three particles
are identified.
A genuine three particle scattering computation involves a wide array of
kinematical invariants and therefore allows a detailed comparison of
the tensorial structures of amplitudes in the two theories. 
Thanks to the absence of lower order ${\cal R}^2$ 
or ${\cal R}^3$ couplings, the correction to the eikonal three
graviton scattering in $d=11$ supergravity induced
by the M-theoretic ${\cal R}^4$ term 
is easily computed via Feynman diagrams and takes a rather simple form, 
as we will see in what follows.

We should remark that this question has been studied before in 
the context of two
graviton scattering~\cite{KV} where it was found that while the
scaling dependence in $v$, $r$, the Planck mass $M$ and the compactified 
radius $R$ is indeed correct, there is however a mismatch of factors of $N$.
In principle one may be content with this mismatch but 
a number of questions remain open. In particular one might 
think that the simple introduction of
the factors of $N$ in what really amounts to an $N=2$ calculation 
in~\cite{KV} is somewhat naive
since it does not take into account bound state effects.
This is reflected in the fact that we have no
control over the Matrix Theory LSZ procedure for two or
three particle scattering for arbitrary values of $N$,
essentially because the ground-state Matrix Theory wavefunction is
still unknown\footnote{Progress towards understanding at least the
asymptotics of the ground-state wavefunction may be found in~\cite{hal,hop}.}.
{}From the viewpoint of the finite $N$ matrix conjecture of 
Susskind~\cite{suss}, however, we are no longer subject to such a restriction.
Is it then possible to find a stronger and more
conclusive test? We believe that a detailed comparison of tensorial 
structures of the three graviton amplitudes of the two models provides
such a test. In addition the abovementioned precise and complete 
agreement found for $3\rightarrow 3$
graviton scattering at leading order to be presented, gives one great
confidence in our methods.

The results of our analysis show a definitive 
disagreement between the 
next to leading Matrix Theory and quantum corrected supergravity amplitudes. 
We find different
tensorial structures in the amplitudes of the two models, 
thus ruling out the proposed correspondence 
of $v^8$ two loop Matrix Theory and ${\cal R}^4$
corrected supergravity.

The outline of the paper is as follows. We present
our supergravity computation of eikonal three graviton scattering 
at leading order in subsection 2.1 and include the
${\cal R}^4$ correction in subsection 2.2. In section three we turn to 
Matrix Theory, where we compute 
the leading S-matrix contribution
to three-particle scattering; in section four we expand this amplitude to 
obtain the next-to-leading
$v^8$ term. In the conclusions, we present the possible viewpoints
explaining the mismatch we have found.

\section{Three graviton scattering in $d=11$ supergravity}

\subsection{Computation of the leading (tree-level) $S$-matrix}  

By definition M-theory at low energies is eleven dimensional supergravity
\cite{Cremmer}, whose bosonic sector is given by the action
\bea
{\cal L}_{0}&=& -\frac{1}{2\kappa_{11}^2} \sqrt{-g}\, 
{\cal R} -\frac 18 \sqrt{-g}\,
(F_{MNPQ})^2
\nn\\&& -\frac{\sqrt{3}}{12^3\kappa_{11}}\varepsilon^{M_1\ldots M_{11}}
F_{M_1M_2M_3M_4}\, F_{M_5 M_6M_7M_8}\, C_{M_9M_{10}M_{11}} \, ,
\label{supergravity}
\eea
where $F_{MNPQ}=4\,\partial_{[M}C_{NPQ]}$, 
$g={\rm det}\,g_{MN}$ and $M=0,\ldots,10$. $\kappa_{11}$ is the 
eleven dimensional gravitational coupling constant. 
Perturbative quantum gravity may be studied by considering small 
fluctuations $h_{MN}$ around the flat metric $\eta_{MN}$, i.e. 
$
g_{MN}= \eta_{MN}+ \kappa_{11}\, h_{MN}
$.
After employing the harmonic gauge 
$\partial_N \, h^N{}_M-(1/2)\partial_M h^N{}_N=0$, 
one derives the graviton propagator 
\be
\langle h_{MN}(k) h_{PQ}(-k)\rangle = -\frac{i/2}{k^2+i\epsilon}
\Bigl ( \eta_{MP}\,\eta_{NQ}+\eta_{MQ}\,\eta_{NP}- \frac 2 9 \,\eta_{MN}
\,\eta_{PQ}
\Bigr )\, .\label{propagator}
\ee
We want to study three graviton scattering at tree level.
At this order, as can be easily seen from the supergravity 
action~(\ref{supergravity}),
the only contribution comes from the pure gravity sector, that
is the Einstein-Hilbert term. In particular, in our computations
we shall need the three-graviton and four-graviton vertices
arising from its expansion.
These are rather lengthy expressions and may be 
found in~\cite{sannan}.

We consider now the elastic scattering process 
$1 + 2 +3 \rightarrow 1'+2'+3'$
of three gravitons into three gravitons and concentrate only on the
terms in the amplitude proportional to
\be
({h_1}\cdot {h_1'})\, ({h_2}\cdot {h_2'})\, 
({h_3}\cdot {h_3'})\, ,
\label{polar}
\ee
${h_i}$ being the external transverse graviton polarization tensors and 
$({h_1}\cdot {h_1'})\equiv h_1^{mn}{h'}_1^{mn}$. The eleven 
dimensional momenta are conveniently parametrized in a light-cone
frame $M=(+,-,m)$ as 
\be
p_i = \Bigl [ -\frac{1}{2} (\v{i}-\frac{\q{i}}{2} )^2 \, ,\, 
1 \, , \, \v{i}-\frac{\q{i}}{2}
\Bigr ]\; , \;\;\;\;
p_i' = \Bigl [ -\frac{1}{2} (\v{i}+\frac{\q{i}}{2} )^2 \, ,\,  
1 \, ,\,  \v{i}+\frac{\q{i}}{2}
\Bigr ] \label{kinematics}
\ee
where $p_i^2=0={p_i'}^2$ and $i=1,2,3$, using a vector notation
for the $SO(9)$ indices $m=1,\ldots , 9$. 
Note that we are considering only
processes with zero compactified $q_-$ 
momentum transfer between in-going particles $i$ 
and outgoing ones $i'$. Conservation of transverse momentum and energy implies
\be
\q{1}+\q{2}+\q{3}=0\, , \qquad
\v{1}\cdot\q{1} + \v{2}\cdot\q{2} + \v{3}\cdot\q{3}=0\, .
\label{emconservation}
\ee
Moreover we will study the amplitude in an eikonal limit.
To be precise this means we keep only terms with at least a double
pole ($1/(q_{1}^2 q_{2}^2)$ and permutations). Terms in which this
minimal pole structure is cancelled represent contact interactions
and cannot be reliably computed in 
the eikonal Matrix Theory framework we present here.
At tree level 
there are  then only the three types of diagrams of figure 1 up to permutations
of the external legs.

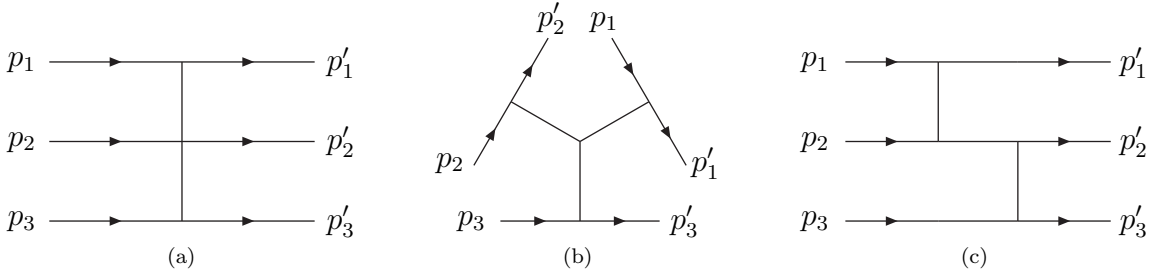
\begin{figure}[tp]
\begin{center}
\begin{picture}(400,100)
\ArrowLine(0,60)(50,60)\ArrowLine(50,60)(100,60)
\Text(-5,60)[rc]{$p_1$}\Text(105,60)[lc]{$p_1'$}
\ArrowLine(0,30)(50,30)\ArrowLine(50,30)(100,30)
\Text(-5,30)[rc]{$p_2$}\Text(105,30)[lc]{$p_2'$}
\ArrowLine(0,0)(50,0)\ArrowLine(50,0)(100,0)
\Text(-5,0)[rc]{$p_3$}\Text(105,0)[lc]{$p_3'$}
\Line(50,60)(50,0)
\Text(50,-10)[ct]{\scriptsize (a)}

\ArrowLine(170,0)(200,0)\ArrowLine(200,0)(230,0)
\Text(165,0)[rc]{$p_3$}\Text(235,0)[lc]{$p_3'$}
\Line(200,0)(200,30)
\Line(200,30)(174,45)\Line(200,30)(226,45)
\ArrowLine(174,45)(188,69)\ArrowLine(160,21)(174,45)
\Text(157,22)[rc]{$p_2$}\Text(190,74)[cb]{$p_2'$}
\ArrowLine(226,45)(240,21)\ArrowLine(212,69)(226,45)
\Text(243,22)[lc]{$p_1'$}\Text(210,74)[cb]{$p_1$}
\Text(200,-10)[ct]{\scriptsize (b)}

\ArrowLine(300,60)(335,60)\Line(335,60)(365,60)\ArrowLine(365,60)(400,60)
\Text(295,60)[rc]{$p_1$}\Text(405,60)[lc]{$p_1'$}
\ArrowLine(300,30)(335,30)\Line(335,30)(365,30)\ArrowLine(365,30)(400,30)
\Text(295,30)[rc]{$p_2$}\Text(405,30)[lc]{$p_2'$}
\ArrowLine(300,0)(335,0)\Line(335,0)(365,0)\ArrowLine(365,0)(400,0)
\Text(295,0)[rc]{$p_3$}\Text(405,0)[lc]{$p_3'$}
\Line(335,60)(335,30)\Line(365,30)(365,0)
\Text(350,-10)[ct]{\scriptsize (c)}
\end{picture}
\end{center}
\caption{The Einstein-Hilbert graphs, (a) V-type, (b) Y-type and (c) 
``re-scattering'' graphs.}
\label{EH_graphs}
\end{figure}

The straightforward but tedious evaluation of these graphs was
performed with the help of the computer algebra system {\it Form} \cite{Jos}.
There are three diagrams of V-type (a) yielding
\be
A_V= 
2 \,
\frac{q_{1}^2 + q_{2}^2+q_{3}^2}{q_1^2 q_2^2 q_3^2} \, v_{12}^2
 v_{23}^2 v_{31}^2 + {\cal O}(v^5\, q^{-3})\, ,
\ee
where we suppress the terms of higher order in $\v{i}$ and lower order
in $\q{i}$. Similarly, there is only one Y-type graph (b) that can be written
as follows:
\be
A_Y=-
\frac {1}{\,q_1^2 q_2^2 q_3^2} \left[ 
(q_1^2 + q_2^2+q_3^2)\, v_{12}^2
 v_{23}^2 v_{31}^2 
-\Upsilon^2 
\right] + {\cal O}(v^5\, q^{-3})\, ,
\label{Y-type}
\ee
where 
\be
\label{Upsilon}
\Upsilon=\Big(
v_{23}^2 \,\q{2}\cdot \v{12}
 +v_{31}^2 \, \q{3}\cdot \v{23}\,
 + v_{12}^2 \, \q{1}\cdot \v{31}
\Big)\, .
\ee
Notice that the combination $\Upsilon\rightarrow {\rm sgn}(\pi)\, \Upsilon$
under any permutation $\pi$ of the labels $1$, $2$ and $3$.
In particular it is then invariant for cyclic permutations of the three labels.
Finally we have the contributions of the six re-scattering graphs (c):
\bea
A_r&\!=\!&-\frac {1}{\,q_1^2 q_2^2 q_3^2} \,\Bigl\{
\,(q_1^2 + q_2^2+q_3^2)\,v_{12}^2v_{23}^2 v_{31}^2 \,-\,
\Big[\,\Big(\frac{q_1^2v_{12}^2v_{31}^2}{\q{2}\cdot\v{12}}\Big)\,\Upsilon\,+\,
\frac{1}{8}\,
\Big(\frac{q_1^2v_{12}^2v_{31}^2}{\q{2}\cdot\v{12}}\Big)^2
+\,\mbox{cyclic}\
\,\Big]\Bigl\}\nn\\&&\hspace{3cm}
+ {\cal O}(v^5\, q^{-3})\; ,
\eea
where $\mbox{cyclic}$ indicates the two cyclic permutations of the
labels $1$, $2$ and $3$.
Summing these three diagrams up one obtains the final result for
the eikonal three graviton amplitude\footnote{Throughout this paper,
we discard the overall coefficients of complete amplitudes.}:
\be
{\cal A}_{EH}= 
\frac{1}{q_1^2q_2^2q_3^2}
\Bigl\{\,\Upsilon^2
\,+\,\Bigl[\,
\Big(\frac{q_1^2v_{12}^2v_{31}^2}{\q{2}\cdot\v{12}}\Big)\,\Upsilon
\,+\,
\frac{1}{8}\,
\Big(\frac{q_1^2v_{12}^2v_{31}^2}{\q{2}\cdot\v{12}}\Big)^2
\,+\,\mbox{cyclic}\,\Bigl]
\Bigl\}
\,+\, {\cal O}(v^5\, q^{-3})\; .\nn\\
\label{EHressupergravity}
\ee
As discussed in the introduction, we deliberately
omitted the $N$-dependence in the formulae above, because
we have complete control of our LSZ matrix theory procedure for 
$N=1$ only. Anyway they can be easily reintroduced with the net result
that (\ref{EHressupergravity}) takes an overall factor of $N_1N_2N_3$, where
$N_i$ is the $p_-$ momentum of the graviton $i$ and where we normalize
each external leg with a factor of $1/\sqrt{N_i}$.

\subsection{The $t_8\, t_8 \, R^4$ contribution}

\label{R4}

In this subsection we will compute the leading correction, in a small
velocity and momentum transfer expansion, to the eikonal three graviton 
scattering involving the higher derivative ${\cal R}^4$ term.

It has been conjectured in \cite{gv97} that the eleven dimensional supergravity
action should contain a ${\cal R}^4$ term, whose form in uncompactified eleven
dimensions is
\be
S_{{\cal R}^4}=\,\frac{\pi^2}{9\cdot 2^7\kappa_{11}^{2/3}}\int\! d^{11}x\sqrt{-g}\,
t_8\,t_8\,{\cal R}^4
\label{R4S}
\ee
where
\be
t_8\,t_8\,{\cal R}^4=t_8^{M_1 M_2 \ldots M_8}\, t_8^{N_1 N_2 \ldots N_8}
\, {\cal R}_{M_1M_2N_1N_2}\,  {\cal R}_{M_3M_4N_3N_4} 
\, {\cal R}_{M_5M_6N_5N_6}\,  {\cal R}_{M_7M_8N_7N_8} \; .
\label{t8}
\ee

\begin{figure}[tp]
\begin{center}
\begin{picture}(100,100)
\ArrowLine(0,60)(50,60)\ArrowLine(50,60)(100,60)
\Text(-5,60)[rc]{$p_1$}\Text(105,60)[lc]{$p_1'$}
\ArrowLine(0,30)(50,30)\ArrowLine(50,30)(100,30)
\Text(-5,30)[rc]{$p_2$}\Text(105,30)[lc]{$p_2'$}
\ArrowLine(0,0)(50,0)\ArrowLine(50,0)(100,0)
\Text(-5,0)[rc]{$p_3$}\Text(105,0)[lc]{$p_3'$}
\Line(50,60)(50,0)
\Vertex(50,30){3}
\end{picture}
\end{center}
\caption{The ${\cal R}^4$ graph with the ${\cal R}^4$ vertex inserted in the
middle.}
\label{R4_graph}
\end{figure}
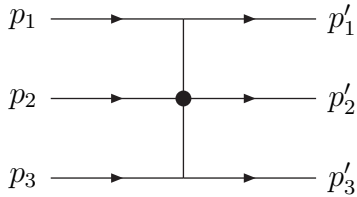

The explicit form of the eight tensor $t_8$ is given, e.g., in \cite{tse8} 
for the 
ten-dimensional case. The tensor $t_8$ entering in (\ref{R4S}),(\ref{t8}),
is obtained by trivially extending the range of the indices to include the
eleventh coordinate. From a supergravity point of view the (linearized)
couplings in (\ref{R4S}) arise as counter terms coming from a one loop four 
graviton 
scattering~\cite{ggv97,rts}. 
In this respect the coefficient in (\ref{R4S}) would be UV 
divergent,
but its finite value is fixed by requiring consistency with results obtained 
in IIA and IIB
string theory \cite{ggv97}.
Explicit computations have excluded the presence of one loop counter terms
of the form ${\cal R}^2$ or ${\cal R}^3$ in $d=11$ 
supergravity\footnote{Strictly
speaking what has been computed in the literature is the  {\it
background} effective action with background fields {\it on-shell} in which
case the absence of ${\cal R}^2$ and ${\cal R}^3$ curvature
corrections has
 been
explicitly verified by Fradkin and Tseytlin~\cite{tsey1}. However it
is an old result~\cite{de_witt} that the $S$-matrix
may be obtained from the on-shell background effective action
by substitution of an iterative solution to the full field equations
of the form $g_{MN}^{\rm cl}=g_{MN}^{\rm as}+\cdots$ where
$g_{MN}^{\rm as}$ is an asymptotic field on mass shell depending
physical polarizations (in particular, here we must take $g_{MN}^{\rm as}$
to be an asymptotic scattering solution in a flat background). 
In practice, this amounts to adding all
possible trees to the effective vertices given by~\eqn{R4S}.}.
It is then not difficult to realize that the first leading
contribution 
to the eikonal 
three graviton scattering involving the couplings (\ref{R4S}), 
is the unique graph
shown in figure 2, that involves the linearized piece of each of the four
Riemann tensors appearing in (\ref{t8}). Any other possible
contribution, 
involving
for instance Y-type or re-scattering-type graphs will be either
sub-dominant 
in a small 
velocity and momentum transfer expansion or outside the eikonal
kinematical 
regime.
We then need to compute only one tree level graph with the insertion
of the ${\cal R}^4$ term as shown in figure 2 
(up to permutations of the external legs).
This can be most easily done by noticing
that the linearized tensorial structure appearing in (\ref{t8}) is 
precisely the same
as that obtained by computing four graviton tree level scattering in a theory
of pure gravity in any space-time dimension \cite{sannan} 
\footnote{It should be noted that in this fashion one only obtains
the {\it on-shell} vertex function. The key observation is that
in the eikonal and spin-less limit (where one discards
terms cancelling the double pole as well as 
contractions of momenta with polarizations) the two {\it a priori}
off-shell legs entering the ${\cal R}^4$ vertex are effectively put on-shell.}.
By using the results
of \cite{sannan} the computation of the graph in figure 2 is then
greatly simplified.
We find that the result for the part of the amplitude
with the external polarizations contracted as in (\ref{polar}) and in the
kinematical parameterization (\ref{kinematics}), can be written as
follows (neglecting an overall coefficient):
\be
{\cal A}_{{\cal R}^4}=\left\{\frac{1}{q_1^2\,q_3^2} \left[
v_{12}^2\,v_{23}^2\,q_2^2+ 
\Upsilon\,(\q{1}\cdot\v{12})\right]^2 
+ \;\; {\rm cyclic}\, \right\}\; ,
\label{R4supergravity}
\ee
where $\Upsilon$ was defined in~\eqn{Upsilon}.
A clarification is now needed. The result (\ref{R4S}), from which we 
computed the graph in figure 2 using the kinematics~(\ref{kinematics}),
applies strictly to  eleven uncompactified space-time dimensions.
However, the correspondence with Matrix Theory
at finite N requires a compactification on an almost time-like 
circle~\cite{suss}.
This means that we should have first compactified the theory on a
spacelike circle and then
performed a computation analogous to that reported in \cite{ggv97}, e.g.
a one loop four graviton scattering with two of them, according to figure 2, 
carrying equal and non-vanishing Kaluza Klein momentum. 
This would give the 
counter term of the form (\ref{R4S}) which has the correct compactified radius 
$R$ and Planck constant $\kappa_{11}$ dependence to match the two loop
Matrix Theory computation we consider in this article, but also terms with 
inappropriate $R$ dependence, namely the analogs of the $\zeta(3)/R^3$
found in~\cite{ggv97} for the case of four graviton scattering with all
external legs carrying vanishing Kaluza Klein momentum. To reach the
discrete light cone kinematics of~\eqn{kinematics} one must take the
limit $R\rightarrow 0$, so that such additional terms should, in
principle, not be neglected. However, our philosophy is to study only
those terms having the right dependence in the radius $R$ and Planck 
constant $\kappa_{11}$ to match two-loops in Matrix Theory perturbation theory.
In particular, (\ref{R4supergravity}) does not  represent the complete eikonal,
leading ${\cal R}^4$ correction at the three graviton scattering for $d=11$ 
supergravity on a circle,
but only the terms that have a chance to be reproduced by a
perturbative two loop Matrix Theory
computation involving supergravitons.
The $N$-dependence of (\ref{R4supergravity}), that we omitted,
is easily computed to be globally of order $N^5$, in disagreement
with the $N^3$ dependence arising at two loops in Matrix Theory.
This reproduces indeed the disagreement found in \cite{KV}.

\section{Scattering gravitons in Matrix Theory}

\label{Matrix}

We now turn to the two loop Matrix Theory calculation, which 
has been carefully computed to leading order 
by Okawa and Yoneya \cite{oy}. 
We have reconsidered their computation
and find results in accordance with theirs.
Importantly, however, we rectify a hole in the original supergravity--
Matrix Theory
agreement presented in \cite{oy}. In more detail, the
technical assumption made by \OaY that all inner products of
impact parameters $\b{ij}$ and relative velocities $\v{ij}$ vanish,
$\{\b{}\cdot \v{}\}=0$, can be shown to pose no restriction for one and two
particle dynamics. But for three particles it constitutes a genuine
restriction. We will show that this restriction may be dropped
rather easily in our framework of comparing S-matrices.

\subsection{The Setup}

Let us summarize the Okawa--Yoneya result in our
notation. The Euclidean Matrix Theory action reads
(setting the Yang--Mills coupling and compactified radius to unity)
\be
\label{e:matrix_action}
S=\int\! dt\,\tr\, \Bigl [ \frac 1 2\,  (D_t X^m)^2 
- \frac 1 4  [X^m,X^n]^2
+ \frac 1 2  ( \psi^T D_t \psi - \psi^T \gamma_m [X^m,\psi] ) \Bigr ],
\ee
where $D_t X^m=\partial_t\,X^m-i\,[A,X^m]$ and $D_t \psi=
\partial_t\psi-i\,[A,\psi]$;
$A$, $X^m$ and $\psi_\alpha$ are hermitian $N\times N$ matrices, 
($m=1,\ldots,9$
and $\alpha=1,\ldots, 16$). Moreover we employ a real symmetric 
representation for the Dirac matrices $\gamma_m$ in which the 
charge conjugation matrix $\cal{C}$ equals unity. 
The background field effective action is computed as an expansion
of the bosonic matrices $X^m_{ij}$ around diagonal backgrounds
\be
X^m_{ij}=\delta_{ij}(b_i^m\,+\,v_i^m\, t) + Y^m_{ij}\, , \qquad
i,j=1,\ldots,  N \; ,
\ee
with constant velocities $\v{i}=v_i^m$, impact parameters $\b{i}=b_i^m$
and fluctuations $Y^m_{ij}$.
As we will focus on the leading spin-independent terms 
in scattering amplitudes, we do not consider fermionic background fields.
Manifestly, this background solves the classical
equations of motion. Thanks to the decoupling of the free $U(1)$
center of mass sector of the model, all one and higher loop results may be
expressed in terms of the relative quantities
$\v{ij}\equiv \v{i}-\v{j}$ and $\r{ij}(t)=\b{ij}+\v{ij}t$.

One proceeds by fixing a background field gauge and adding
appropriate ghost couplings and kinetic terms. 
The propagators for all
fluctuations may be expressed in terms of the inverse of a kinetic
operator $-\partial_t^2+r_{ij}(t)^2$ which, in proper time
representation, reads
\be
\Big[-\partial_t^2+r_{ij}(t)^2\Big]^{-1}\circ\, \delta(t_1-t_2)\,=\,
\int_0^\infty\! d\s\,\,
\Delta\Big(\s,t_-,\r{ij}(t_+ )\Big)\; ,
\ee
where $t_-=(t_1-t_2)/2$ and $t_+=(t_1+t_2)/2$ 
along with
\bea
\Delta\Big(\s,t_-,\r{ij}(t_+)\Big)&=&
\sqrt{\frac{v_{ij}}{2\pi\sinh(2\o v_{ij})}}\label{prop}\,\exp\Big[
-v_{ij}^2 t_-^2 \coth(\o v_{ij}) -\o
r_{ij}(t_+)^2 \nn \\
&&-\Big(\frac{\v{ij}\cdot
\r{ij}(t_+)}{v_{ij}}\Big)^2\frac{1}{v_{ij}}\Bigl(\tanh(\o v_{ij})-\o v_{ij}
\Bigr)
\Big]\; ,
\eea
using $v_{ij}=|\v{ij}|$.
The two loop calculation is then rather standard, yet tedious.
One computes the  
three and four point vertices from the expansion of the
action \eq{e:matrix_action} about the background. There are three
possible topologies, the dumbbell, setting sun and figure eight 
denoted $\Gamma_{\rm o\!-\!o}$, $\Gamma_Y$ and $\Gamma_V$, 
respectively\footnote{To be precise, note that any terms from the 
setting sun diagram that may be written as a total derivative $d/do_i$ 
of a polynomial times three propagators are included in $\Gamma_V$ 
rather than $\Gamma_Y$, see~\cite{oy} for details.}
as depicted in figure~3 in 't Hooft double line notation. 
We remark that, as can clearly be
seen from the Matrix Theory LSZ formalism formulated
in~\cite{psw,psw1,psw2}, one must compute {\it all} Matrix Theory graphs,
one particle irreducible, connected-reducible and 
disconnected\footnote{This is easily seen as follows. In quantum 
mechanics the $S$-matrix reads $S_{fi}=\int dx'\, dx\,$ 
$\Phi_f^*(x')\,\langle x'| \exp(-iHT)|x\rangle \Phi_i(x)$ for incoming 
and outgoing wavefunctions $\Phi_i$ and $\Phi_f^*$. The transition element 
from $|x\rangle$ to $\langle x'|$ may be represented as a path integral 
for which clearly one must compute all diagrams.}.
The latter we
disregard since it is easy to see that they can only correspond to
disconnected graphs on the supergravity side. 
However, as we shall see,
graphs of the connected-reducible type (such as the dumbbell graph)  
reproduce re-scattering processes in supergravity \cite{oy1}.

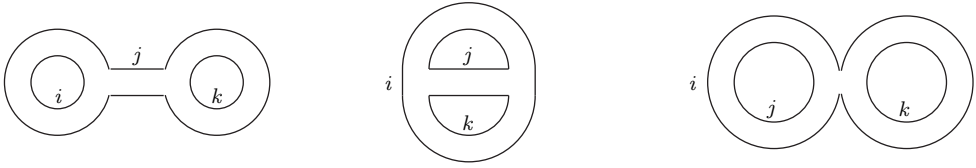
\begin{figure}[tp]
\begin{center}
\begin{picture}(350,70)
\Line(20,35)(40,35)\Line(20,25)(40,25)
\CArc(0,30)(20,14,346)\BCirc(0,30){10}
\CArc(60,30)(20,194,166)\BCirc(60,30){10}
\Text(30,40)[cc]{\scriptsize \it j}
\Text(0,25)[cc]{\scriptsize \it i}
\Text(60,25)[cc]{\scriptsize \it k}

\CArc(270,30)(25,10,350)\BCirc(270,30){15}
\CArc(320,30)(25,190,170)\BCirc(320,30){15}
\Text(270,20)[cc]{\scriptsize \it j}
\Text(320,20)[cc]{\scriptsize \it k}
\Text(240,30)[cc]{\scriptsize \it i}

\Line(140,35)(170,35)\Line(140,25)(170,25)
\CArc(155,35)(15,0,180)\CArc(155,25)(15,180,360)
\CArc(155,35)(25,0,180)\CArc(155,25)(25,180,360)
\Line(180,35)(180,25)\Line(130,35)(130,25)
\Text(155,40)[cc]{\scriptsize \it j}
\Text(155,15)[cc]{\scriptsize \it k}
\Text(125,30)[cc]{\scriptsize \it i}

\label{MT_graphs}
\end{picture}
\end{center}
\caption{The planar 2 loop Matrix Theory graphs: dumbbell, setting sun and 
figure eight.}
\end{figure}

The \OY result may be stated (somewhat implicitly) 
as the effective action
\be
\label{e:OY_result}
\Gamma_{2\rm\, loop}=\Gamma_{\rm o\!-\!o}+\Gamma_V+\Gamma_Y
\ee 
where 
\be
\label{e:dumbbell}
\Gamma_{\rm o\!-\!o}=-\frac{1}{2}\,\sum_{i}\,\int\! dt_1dt_2\,
\langle\, \partial_{t_1}^2 Y_{ii}^m(t_1)\,\rangle\,
\Delta(t_1-t_2)
\langle\, \partial_{t_2}^2 Y_{ii}^m(t_2)\,\rangle\; ,
\ee
with
\bea
\langle\, \partial_{t}^2 Y_{ii}^m(t)\,\rangle&=&
-32\,\sum_j\int_0^\infty\! d\o\, \label{e:tadpole}\\&&\quad
\Big(
r^m_{ij}(t)\,\sinh^4(\frac{\o v_{ij}}{2})
+\frac{v^m_{ij}}{v_{ij}}\,\cosh(\frac{\o v_{ij}}{2})\,
\sinh^3(\frac{\o v_{ij}}{2})
\frac{\partial}{\partial t}\,
\Big)\,
\Delta(\o,0,\r{ij}(t)) \; ,\nn
\eea
and
$\Delta(t_1-t_2)=\int_0^\infty\! d\o\Delta(\o,t_-,0)$ is the propagator
for a free massless scalar field in one dimension. Further
\bea
\label{e:Gamma_V}
\Gamma_V&=&-128 
\,\sum_{ijk}\int \!
dt \int_0^\infty\! d\o_1 d\o_2\,
\sinh^3(\frac{\o_1 v_{ij}}{2})\,
\sinh^3(\frac{\o_2 v_{jk}}{2})\,\nn\\&&
\Big(
\frac{2\,\v{ij}\cdot \v{jk}}{v_{ij}v_{jk}}\,
\cosh(\frac{\o_1 v_{ij}}{2})\,
\cosh(\frac{\o_2 v_{jk}}{2})\,-\,
\sinh(\frac{\o_1 v_{ij}}{2})\,
\sinh(\frac{\o_2 v_{jk}}{2})\,
\Big)\nn\\&&
\Delta(\o_1,0,\r{ij}(t))\,\Delta(\o_2,0,\r{jk}(t))
 \; ,
\eea
along with
\bea
\Gamma_Y&=&- 
\sum_{ijk}\int\! 
dt_+ dt_- \int_0^\infty\! d\o_1 d\o_2 d\o_3\, 
P_Y(\o_1,\o_2,\o_3,\r{ij}(t_+),\r{jk}(t_+),\r{ki}(t_+) 
\v{ij},\v{jk},\v{ki},t_-)\nn\\&& 
\hspace{3cm} \Delta(\o_1,t_-,\r{ij}(t_+))\,\Delta(\o_2,t_-,\r{jk}(t_+))
\,\Delta(\o_3,t_-,\r{ki}(t_+)) \label{e:Gamma_Y} \; .
\eea
The \OY computation of the function $P_Y$ is an impressive technical 
achievement and the result is a quadratic polynomial in the variables
$\r{ij}(t_+)$ and $t_-$ (the result itself is given by equation (3.47) 
of~\cite{oy} along with three pages of the appendices of that work).
Its correctness (at least to leading order in $\v{ij}$)
is well tested by comparison with supergravity. 

A remark on the $N$ dependence of the two loop effective action 
$\Gamma_{\rm 2\,loop}$ is in order. The planar two
loop graphs of figure~3 carry three independent $U(N)$
indices ($i,j,k$) thus giving rise to three body interactions.
For backgrounds consisting of
three blocks proportional to unit matrices of size $N_i$ ($i=1,2,3$,
with $\sum_i N_i=N$) the sums $\sum_{ijk}$ reduce to
$N_1 N_2 N_3\, \sum_{ijk=1}^3$ and $\Gamma_{\rm 2\,loop}$ scales
homogeneously like $N^3$ to {\it all} orders in $\v{ij}$, precisely
like the corresponding supergravity term (\ref{EHressupergravity}).
This procedure, however, has from our viewpoint no real justification
and we will therefore take $N_i=1$ in the following.

Up to now we have simply restated the results of~\cite{oy}. 
In what follows
we compare these results with the tree level supergravity $S$-matrix and
in doing so show how to relax the restriction $\{\b{}\cdot \v{}\}=0$. 
Thereafter, the same techniques will be employed to compare 
the next to leading order in $\v{ij}$ Matrix Theory prediction 
with one loop supergravity.

\subsection{$\Gamma_Y$ contribution to the Matrix Theory $S$-matrix}

Let us begin with the most difficult contribution $\Gamma_Y$ of
\eq{e:Gamma_Y}. One might suspect that since the result depends 
on three proper time
parameters $\o_1$, $\o_2$ and $\o_3$ the result ought correspond to the
triple pole structure of the Y-type diagrams in supergravity 
and indeed this naive suspicion will be
borne out in the following. According to the Matrix Theory LSZ
formalism~\cite{psw,psw1,psw2} the leading spin
independent 
$1+2+3\rightarrow 1'+2'+3'$ Matrix Theory $S$-matrix is given by 
\be
\label{eq:S-matrix}
S^{3\rightarrow3}=
\int\! d^9b_1 d^9b_2 d^9b_3 \, \exp(iq_1\cdot b_1+iq_2\cdot
b_2+iq_3\cdot b_3)\,\Gamma_{\rm 2\,loop} \, .
\ee
Note that we have dropped contributions
corresponding to disconnected processes (so that 
$\Gamma_{\rm 2\,loop}$ no longer appears in the exponent). 
The transverse kinematics described
by~\eqn{eq:S-matrix} are initial and final momenta
\bea
{\bf p}_i&=&\v{i}-\q{i}/2\nn\\  
{\bf p^\prime}_i&=&\v{i}+\q{i}/2\; ,\qquad i=1,2,3 \; ,
\eea  
in accord with the supergravity kinematics \eq{kinematics}.
Note that at this stage $\v{i}$ is {\it not} a velocity anymore,
but rather the average momentum of the $i$'th particle
$\v{i}=({\bf p}_i+{\bf p}_i')/2$, for details see \cite{psw,psw1,psw2}.
Since $\Gamma_{\rm 2\,loop}$ only depends on relative quantities,
the integral over the average impact parameter 
$(\b{1}+\b{2}+\b{3})/3$ yields the usual
momentum conserving
$\delta^{(9)}(\q{1}+\q{2}+\q{3})$ which we drop from now on. 
Concentrating on the $\Gamma_Y$ contribution we then have
\bea\label{311}
S_Y^{3\rightarrow3}&=& -
\int\! d^9b_{13}\, d^9b_{23} \, 
\exp(iq_1\cdot b_{13}+iq_2\cdot
b_{23})\, 
\int\! 
dt_+ dt_- \int_0^\infty\! d^3\o \\
&& P_Y(\o_i,\r{ij}(t_+),\v{ij},t_-)
\Delta(\o_1,t_-,\r{12}(t_+))\,\Delta(\o_2,t_-,\r{23}(t_+))
\,\Delta(\o_3,t_-,\r{31}(t_+)) \; .\nn
\eea 
The leading contribution to three body scattering should depend on the
sixth power of velocities $\v{12}$, $\v{23}$ and $\v{31}$ as can be seen from
the supergravity amplitude \eq{EHressupergravity}. However, if one
examines the polynomial $P_Y$, its leading behavior is quadratic in 
velocities and the ``propagators'' $\Delta$ are to leading order
velocity independent. In order to see explicitly how the cancellations of the
terms quadratic and quartic in velocities occur, 
two observations are needed. Firstly, examining 
the $t_-$ dependence of the
exponent in \eq{311} arising from the three propagators $\Delta$
defined in \eq{prop} 
\be
-t_-^2(v_{12}^2\coth(\o_1 v_{12})+v_{23}^2\coth(\o_2 v_{23})+
v_{31}^2\coth(\o_3 v_{31}))\, \equiv \, -t_-^2\,P \; ,
\ee 
one sees that under the Gaussian $t_-$ integral, all terms linear in $t_-$
can be discarded by symmetric integration and terms proportional to
$t_-^2$ may be replaced by $1/(2P)$. Secondly, observe that the
operator $d/dt_+$ acting on the three propagators $\Delta$ in
\eq{311} yields the factor
\be
-2\,\Big[
\frac{\v{12}\cdot \r{12}(t_+)}{v_{12}}\tanh(\o_1 v_{12})+
\frac{\v{23}\cdot \r{23}(t_+)}{v_{23}}\tanh(\o_2 v_{23})+
\frac{\v{31}\cdot \r{31}(t_+)}{v_{31}}\tanh(\o_3 v_{31})
\Big ]
\,\equiv\,Q\, .
\ee
Now recall that $P_Y$ is a polynomial quadratic in $\r{12}(t_+)$, 
$\r{23}(t_+)$ and $\r{31}(t_+)$. However, intuitively one may expect
that terms of order two and four in velocity should not depend on the
impact parameters $\b{i}$ since, in the case of one and two particle 
kinematics, shifts of the zero of $t_+$ can always be
made in such a fashion as to arrange that
$\r{ij}(t_+)\rightarrow \v{ij} t_+$. This in fact is the case since at
orders two and four in velocity, the $\r{ij}$ dependence of $P_Y$ can be
expressed as $Q\times$(terms order one in velocity). Writing Q as
$d/dt_+$ acting on the $\Delta$'s and subsequently integrating by
parts removes all dependence on $\r{12}$, $\r{23}$ and $\r{31}$.
Coupled with the first observation, one in fact finds
miraculously that all terms proportional to squares and the fourth
power of velocity cancel \cite{oy}. We stress that no
restriction involving inner products of velocities and impact
parameters must be imposed for this cancellation to take place. 

It is now advantageous to interchange the $dt_+$ and $d^9\b{}$ integrals
and thereafter shift the integration variable $\b{13}\rightarrow 
\r{13}(t_+)$ along with $\b{23}\rightarrow \r{23}(t_+)$ so that 
the $t_+$ integral may
be performed yielding an energy conserving delta function
\bea
S_Y^{3\rightarrow3}&=&- 
\,(2\pi)\delta(\q{1}\cdot \v{13}+\q{2}\cdot\v{23})\,\int\! d^9\r{13} d^9\r{23} \, 
\exp(i\q{1}\cdot \r{13}+i\q{2}\cdot
\r{23})\, \int\!dt_- \int_0^\infty\! d^3\o\nn \\&& 
\widetilde P_Y(\o_i,\r{ij},\v{ij})\,
\Delta(\o_1,t_-,\r{12})\,\Delta(\o_2,t_-,\r{23})
\,\Delta(\o_3,t_-,\r{31})\; , 
\eea
where the tilde over $P_Y$ indicates that we have performed 
the manipulations indicated in the two observations above. 

So far we have managed to rewrite the $\Gamma_Y$ contribution to the Matrix
Theory $S$-matrix as (suppressing from now on 
the energy conserving delta function
$(2\pi)\delta(\q{1}\cdot \v{13}+\q{2}\cdot\v{23})$) 
\bea
\label{e:S/2}
S^{3\rightarrow3}_Y&=&-
\,\int\! d^9\r{13} d^9\r{23} \, 
\exp(i\q 1\cdot \r{13}+i\q 2\cdot
\r{23})\,\int_0^\infty \!d^3\o\,\frac{1}{\sqrt{4\pi P}}  \\
&& (\widetilde P_Y+ \widetilde P_Y{}^m r^m +r^m\widetilde P_Y{}^{mn}
r^n)\,
\Delta(\o_1,0,\r{12})\,\Delta(\o_2,0,\r{23})
\,\Delta(\o_3,0,\r{31})\; . \nn
\eea
Note that we have performed the integral over $t_-$ as explained above.
Furthermore, $\widetilde P_Y$, $\widetilde P_Y{}^m$ and 
$\widetilde P_Y{}^{mn}$
are functions of the $\sigma_i$, $\r{ij}$ and $\v{ij}$ only 
($ab=(12,23,31)$) and their leading behavior
goes with the sixth power of velocity. Also their coupling to $\r{ij}$
has been schematized.

We proceed by interchanging the Fourier integrals over $\r{13}$ and
$\r{23}$  with those over proper 
time $\sigma_i$ parameters. If we content 
ourselves with leading order in velocities, the
$\r{}$ dependence in the exponent of~\eqn{e:S/2} reads
\be
\exp
\Big(
 i\q 1\cdot \r{13}+i\q 2\cdot
\r{23}-r_A^m {\cal O}_{AB}r_B^m
\Big)\, ,\qquad
{\cal O}=\left(
\begin{array}{cc}
\o_1+\o_3&-\o_1\\
-\o_1&\o_1+\o_2
\end{array}
\right)\, ,
\ee
where the index $A=(13,23)$. The matrix ${\cal O}$ has determinant 
$p=\o_1\o_2+\o_2\o_3+\o_3\o_1$ and the Gaussian integral over $\r{13}$
and $\r{23}$ may now be performed. Remarkably, we find that all terms not
proportional to inner products of momentum transfers $\q{i}$ and
velocities $\v{ij}$ cancel amongst themselves and to leading order in
velocities we are left with
\be
S^{3\rightarrow3}_Y=\frac{\pi^8}{4}
\,\int_0^\infty
\!d^3\o\,\frac{1}{p^3}\,\exp\Big(-\frac{1}{4p}(q_1^2 \o_2+q_2^2 \o_3 +
q_3^2 \o_1)\Big)\,\Upsilon^2 \; .
\ee
where $\Upsilon$ is the same as defined in (\ref{Upsilon}).
Finally doing the $d^3\o$ integral yields our result
\be
\label{e:S_Y}
S^{3\rightarrow3}_Y=32\,\pi^9\,
\frac{\Upsilon^2}{q_1^2\,q_2^2\,q_3^2}\; .
\ee
Although we leave the orchestration of the two loop leading velocity
Matrix Theory result to the end of this section, we remark that
\eqn{e:S_Y}
already has precisely the correct form to match with tree level 
supergravity graphs of the Y-type \eq{Y-type} in the triple pole sector. 

\subsection{$\Gamma_V$ contribution to the Matrix Theory $S$-matrix}

Compared with the $\Gamma_Y$ contribution, the computation of the
$S$-matrix elements arising from the $\Gamma_V$ terms are very 
straightforward. The leading contribution from $\Gamma_V$ as given
in~\eqn{e:Gamma_V} is seen by inspection to be order six in
velocity. Hence,
interchanging $dt$ and $d^9\b{}$ integrals as above and
thereafter performing the Fourier transforms and proper time $\sigma_i$
integrations we find (suppressing delta functions over energy and momentum)
\be
S^{3\rightarrow3}_V=-64\pi^9 \,
\frac{v_{12}^2\,v_{31}^2\,\v{12}\cdot \v{31}}{q_2^2\,q_3^2}\,
+ \mbox{cyclic}\, .
\label{e:S_V}
\ee
We emphasize that the result~\eqn{e:S_Y} mixes with terms 
arising from dumbbell graphs $\Gamma_{\rm o\!-\!o}$
which we will consider next. Thus
a comparison to supergravity is not possible until we consider
the sum of {\it all} Matrix Theory Feynman diagrams, which has
been the source of some confusion in the literature \cite{dr,dine2}.

\subsection{$\Gamma_{\rm o\!-\!o}$ contribution to the Matrix Theory
$S$-matrix}
\label{recoil}

The final Matrix Theory contribution to the leading order
$3\rightarrow3$ $S$-matrix is given by the dumbbell diagrams.
In~\cite{oy1} it has been shown that these graphs can be
given the interpretation of recoil corrections to a source probe
approximation. In Feynman diagram language there is, of course, no
artificial distinction into recoil and non-recoil terms (physically
since one finds that $\Gamma_V$ and $\Gamma_{\rm o\!-\!o}$
contributions mix, this is certainly the case). 

To extract the $S$-matrix contribution from $\Gamma_{\rm o\!-\!o}$
as given in~\eqn{e:dumbbell} and~\eqn{e:tadpole} we begin by writing
the free massless propagator for a scalar field in one dimension as
\be
\Delta(t_1-t_2)=\int \frac{d\omega}{2\pi} \frac{e^{-i\omega
(t_1-t_2)}}{\omega^2+i\epsilon}\, . 
\ee
The explicit time derivatives appearing in the truncated
tadpoles~\eqn{e:tadpole} may, integrating by parts, be converted to 
$\omega$'s. Then, in the same fashion explained above, interchanging
$d^9\b{}$ and time integrals and shifting $\b{}\rightarrow \r{}(t)$, then
performing the resulting Fourier transforms and proper time integrals
we find
\bea
S^{3\rightarrow 3}_{\rm o\!-\!o}&=&\pi^7\,
\sum_{i\neq j\neq k}\int\! dt_1 dt_2\,\int\! d\omega
\exp\Big(
-iq_j\cdot v_{ji} t_1-i\omega(t_1-t_2)-iq_k\cdot v_{ki} t_2
\Big) \nn\\&& \frac{1}{\omega^2+i\epsilon}\, 
\frac{v_{ij}^2\,v_{ki}^2}{q_{j}^2\,q_{k}^2}\,
\Big [ \q{j}\, v_{ij}^2-4\,\omega\, \v{ij}\Big]\cdot
\,\Big [ \q{k}\, v_{ki}^2-4\,\omega\, \v{ki}\Big ]
\; .
\eea
Note that we have kept only the leading velocity dependence and
discarded terms in the sum over $U(N)$ indices $i$,
$j$ and $k$ in which the inner loop running around each end of the
dumbbell takes the same value since one may convince oneself that these
terms can only correspond to disconnected processes\footnote{Such terms
have been analyzed in a recent preprint \cite{rtz}.}.  
Now, the integral over $t_-=(t_1-t_2)/2$ yields 
$\delta(2\,\omega+q_j\cdot v_{ji}-q_k\cdot v_{ki})$ and the
$t_+=(t_1+t_2)/2$ integral yields the usual energy conserving delta
function which we suppress as usual. The integral over $\omega$ is then
trivial and gives the final result
\be
\label{e:S_dumbbell}
S^{3\rightarrow 3}_{\rm o\!-\!o}\,=\,
4\,\pi^9\,
\Big [ 16\, \frac{v_{12}^2\,v_{31}^2\, \v{12}\cdot\v{31}}{q_2^2\,q_3^2}\,
+8\,\Upsilon\,\frac{v_{12}^2\,v_{31}^2}{q_2^2\,q_3^2\,\q{2}\cdot\v{12}}
\,+\,\frac{v_{12}^4\,v_{31}^4\, q_1^2}
{q_2^2\,q_3^2\,(\q{2}\cdot\v{12})^2} \,+\,\mbox{ cyclic}\,\Big ]\; .
\ee
Observe in particular that here the first term and its permutations
exactly cancels the contribution from $S_V^{3\rightarrow 3}$ 
in \eqn{e:S_V}. Clearly then, one sees that from a physical
viewpoint the split into recoil and non-recoil terms is an artifact of
one's approximation scheme. In a Feynman graph approach, where one
simply computes all terms contributing at a given order in velocity
there is no need to make such a distinction so long as one also
computes all Feynman diagrams on the Matrix Theory side.

Finally, we see that the sum
$S_Y^{3\rightarrow 3}+ S_V^{3\rightarrow 3}+ S_{\rm o\!-\!o}^{3\rightarrow 3}$
as given in equations~\eq{e:S_Y}, \eq{e:S_V} and \eq{e:S_dumbbell} 
reproduces the tree
level supergravity result \eq{EHressupergravity}. No restriction upon impact
parameters or velocities has been made in this comparison and this
result represents the completion of the leading order spin-independent
three graviton scattering problem whose tortuous history may be
followed in the sequence of articles~\cite{dr,iff,Taylor,oy}.

\section{Next to leading order: can Matrix Theory 
see ${\cal R}^4$ corrections?}

Armed with the above clear cut scheme for the computation of Matrix Theory
S-matrix elements and given the precise agreement of the tree level 
supergravity amplitude with the leading Matrix Theory result,
we now turn to the question of whether Matrix Theory is
sensitive to the one loop corrections to the M-Theory effective
action discussed in section \ref{R4}. A simple dimensional analysis 
indicates that the next to leading order contributions
to the two loop Matrix Theory effective action, i.e. the terms
of order $v^8/(r^{18}\, R^7\, M^{24})$, have 
the correct dependence on $v,r$, the eleven-dimensional Planck mass $M$ and 
compactification radius $R$ to match the ${\cal R}^4$ correction of
\eq{R4S} \cite{2gravR4}.

As mentioned in the introduction, 
this question has been already studied 
for two graviton scattering in \cite{KV}, where a mismatch
of factors $N$ between supergravity and Matrix Theory was found.
However,  our philosophy here
is quite different, since we perform an analysis of tensorial structures
in both theories which will allow us to give 
more definite and stronger conclusions.

The setup of the computation is now clear. We simply expand all
terms in the two loop effective action $\Gamma_{2\rm\, loop}$
of \eq{e:OY_result} to order $v^8$ and apply the same manipulations
discussed in the last section to obtain the Matrix Theory amplitudes.

\subsection{Next to leading order results and disagreement}

The order $v^8$ result of the spin independent $1+2+3\rightarrow
1'+2'+3'$ amplitude is again comprised of the three terms
\be
S^{3\rightarrow3}|_{v^8}=S^{3\rightarrow3}_{\rm o\!-\!o}|_{v^8}
+S^{3\rightarrow3}_{V}|_{v^8}+S^{3\rightarrow3}_{Y}|_{v^8}\, .
\ee
Dropping the overall energy and momentum conserving delta
function we find
\bea
S^{3\rightarrow3}_{\rm o\!-\!o}|_{v^8}&=&-\frac{\pi^9}{6}
\Big [ \,\frac{v_{12}^4\, v_{31}^4}{\q{2}\cdot\v{12}}\Big(
\vac{\ft{1}{\o_1^2}}\, \q{3}\cdot\v{12}+
\vac{\ft{1}{\o_2^2}}\, \q{2}\cdot\v{31}\Big )
+ \q{2}\cdot\q{3}\, v_{12}^4\, v_{31}^4\, \Big ( \vac{\ft{1}{\o_1}}
+\vac{\ft{1}{\o_2}}\Big )\nn\\&&
-4 \,\v{12}\cdot\v{31}\, v_{12}^2\,v_{31}^2\, \Big ( \vac{\ft{1}{\o_1^2}}
\, v_{12}^2+\vac{\ft{1}{\o_2^2}}\, v_{31}^2\Big )\nn\\&&
-4\, \q{2}\cdot\v{12}\, v_{12}^2\, v_{31}^2\, (v_{12}^2\, \q{2}\cdot
\v{31}+v_{31}^2\, \q{3}\cdot\v{12})\, \Big ( \vac{\ft{1}{\o_1}}
+\vac{\ft{1}{\o_2}}\Big )\nn\\&&
+16 \, \v{12}\cdot\v{31}\,
(\q{2}\cdot\v{12})^2\, v_{12}^2\, v_{31}^2\,\Big( \vac{\ft{1}{\o_1}}
+\vac{\ft{1}{\o_2}}\Big )+\mbox{ cyclic}\,\Big ]
\label{v8o-o}
\eea
along with
\bea
S^{3\rightarrow3}_{V}|_{v^8}&=&
\frac{\pi^9}{2}\, v_{31}^4\,v_{12}^4\, \vac{\ft{1}{\o_1\,\o_2}}
-\frac{\pi^9}{3}\, \v{12}\cdot\v{31}\,v_{12}^2\,v_{31}^2\, 
\Big [ v_{12}^2\, \vac{\ft{1}{\o_1^2}}+ v_{31}^2\, \vac{\ft{1}{\o_2^2}}
\nn\\&&
-4\, (\q{2}\cdot\v{12})^2\, \Big ( \vac{\ft{1}{\o_1}}+\vac{\ft{1}{\o_2}}
\Big)\Big ] +\mbox{cyclic}\, ,
\label{v8V}
\eea
where we have defined
\be
\vac{f(\o_1,\o_2)}=\int_0^\infty d^2\o\, f(\o_1,\o_2)\, e^{-\o_1\,q_2^2
-\o_2\, q_3^2}
\ee
that is the proper time integrals remain to be performed\footnote{As
a matter of fact all integrals in \eq{v8o-o} and \eq{v8V} are
divergent, but exist in a distributional sense. See for example 
\cite{distributions}; 
one must interpret the logarithm in~\eqn{log} as 
$\log(q^2/\Lambda^2)$
for some momentum scale $\Lambda$ which can only be determined by some 
physical principle.}. We first
note that none of the terms in \eq{v8o-o} and \eq{v8V} displays
a genuine two pole structure $\vac{1}=1/(q_2^2\, q_3^2)$ as found
in the supergravity amplitude \eq{R4supergravity}, such terms, however, will
arise from the $S^{3\rightarrow3}_{Y}|_{v^8}$ contribution to be
studied.

An immediate disagreement arises from the first term of \eq{v8o-o} with
a ``re-scattering pole'' $1/(\q{2}\cdot\v{12})$, whereas on the
supergravity side re-scattering diagrams of the type (c) of figure
1 are absent since there are no ${\cal R}^2$ and ${\cal R}^3$ curvature
corrections 
to the effective M-Theory action, as argued in section \ref{R4}. 
Note also that  $S^{3\rightarrow3}_{Y}|_{v^8}$ does not give
rise to re-scattering poles, as we shall see shortly.
Performing the corresponding $\o$ integrals for this term
in a distributional sense
\be
\label{log}
\int_0^\infty d\o \, \frac{1}{\o^2}\, e^{-\o\, q^2}
=\frac{1}{16}\, q^2\, (\log{q^2}+\gamma-1) \; ,
\ee
where $\gamma$ is the Euler constant, the re-scattering
contributions of $S^{3\rightarrow3}|_{v^8}$ take the form
\be
\frac{v_{12}^4\, v_{31}^4}{\q{2}\cdot\v{12}}\Big(
\frac{q_2^2}{q_3^2}\, \q{3}\cdot\v{12}+
\frac{q_3^2}{q_2^2}\, \q{2}\cdot\v{31}\Big ) + \mbox{log terms}\; .
\ee
Hence it is clear that Matrix Theory produces terms with no counterpart
on the supergravity side. However, taking a conservative viewpoint one
could argue that only the ``truly eikonal'' terms with a double pole 
$1/(q_2^2\, q_3^2)$ structure should be compared on both sides.
A similar phenomenon occurred in the computation of polarization
dependent two graviton scattering amplitudes 
\cite{psw}, where the spin dependent contributions to the 
Matrix Theory amplitude
gave rises to terms cancelling the $1/q^2$ pole and had to
be dropped. 

Taking this viewpoint we would have to conclude that all terms in
\eq{v8o-o} and \eq{v8V} are spurious and we need to go on to the
rather involved computation of $\Gamma_Y$ at order $v^8$.

The outcome of this computation 
is the amplitude (recall that $p=\o_1\o_2+\o_2\o_3+\o_3\o_1$)
\be
\label{why}
S^{3\rightarrow3}_Y|_{v^8}=\int_0^\infty
\!d^3\o\,\frac{1}{p^5}\exp(-q_1^2 \o_2-q_2^2 \o_3 -
q_3^2 \o_1)\Big(\Upsilon^2\,p^2\, \Pi_2 + \Upsilon \,p\, \Pi_1+\Pi_0\Big)
\ee
where $\Upsilon$ was introduced in~\eqn{Upsilon}
and $\Pi_n$ $(n=0,1,2)$ are polynomials
order $7-n$ in the $\sigma$'s and order $n$ in $\q{}\cdot\v{}$'s. 
In particular
\be
\Pi_2=-\frac{8\pi^9}{3}\Big(\, (\v{12}\cdot\q1)^2\,(\o_1+\o_2)(\o_1\o_2)^2
-2\, \v{12}\cdot\q1\,\v{23}\cdot\q2\,\o_1\o_2^3\o_3\Big)\,+\,\mbox{cyclic}
\ee
and
\bea
\Pi_1\!\!&=&\frac{\textstyle 16\pi^9}{\textstyle3}\,\v{12}\cdot\q1\nn\\&&
\Big(\,
(\o_1^2v_{12}^4+\o_2^2v_{23}^4)\,\o_1\o_2\,[\o_1\o_2-2(\o_1+\o_2)\o_3]
+3v_{31}^4\,(\o_1\o_2\o_3)^2
\nn\\&&
+2v_{12}^2v_{23}^2\o_2\o_1(\o_1^3\o_2+\o_2^3\o_3+3\o_1^2\o_2^2
                +\o_1^2\o_2\o_3+\o_1^3\o_3+\o_1\o_2^2\o_3+\o_1\o_2^3)
\nn\\&&
-v_{23}^2v_{31}^2\,\o_2^2(3\o_1^2\o_3^2-\o_2^2\o_3^2-2\o_1^2\o_2\o_3
                +\o_1^2\o_2^2+2\o_2\o_1\o_3^2)
\nn\\&&
-v_{12}^2v_{31}^2\,\o_1^2(2\o_2\o_1\o_3^2+\o_1^2\o_2^2-\o_1^2\o_3^2
                -2\o_1\o_2^2\o_3+3\o_2^2\o_3^2)
\Big)\,+\,\mbox{cyclic}
\eea
along with
\bea
\hspace{-.9cm}\Pi_0&=&
-\frac{\textstyle 8\pi^9}{\textstyle 3}\Big[\,
v_{12}^8\,\o_1^3(-2\o_3^2\o_2^2-\o_1\o_3\o_2^2-\o_1\o_3^2\o_2
-4\o_1^2\o_3\o_2+\o_3^2
\o_1^2+\o_1^2\o_2^2)
\nn\\
\lefteqn{\!\!\!
-4v_{12}^6v_{23}^2\,\o_1^2(3\o_3^2\o_2^3
+\o_1\o_3^2\o_2^2+5\o_1^2\o_3
\o_2^2-\o_1^2\o_3^2\o_2-3\o_1^2\o_2^3-\o_1^3\o_3\o_2
+\o_3^2\o_1^3-2\o_1^3\o_2^2)}
\nn\\
\lefteqn{\!\!\!
-2\,v_{12}^4v_{23}^2v_{31}^2\,\o_1(\o_1\o_3-2\o_3\o_2+\o_1\o_2)
(3\o_3^2\o_1^2
-3\o_3^2\o_2^2+\o_1^3\o_3-5\o_1^2\o_3
\o_2+\o_1^3\o_2+3\o_1^2\o_2^2)}
\nn\\
\lefteqn{\!\!\!
+\,v_{12}^4v_{23}^4\,(2\o_1\o_3\o_2^5-\o_1\o_3^2\o_2^4+11\o_1^2
\o_3\o_2^4+10\o_1^2\o_3^2\o_2^3+\o_1^2\o_2^5+10\o_1^3\o_3^2\o_2^2
+12\o_1^3\o_2^4}
\nn\\&&\hspace{-1.22cm}
+11\o_1^4\o_3\o_2^2
-\o_1^4\o_3^2\o_2+12\o_1^4\o_2^3+2\o_1^5\o_3\o_2
+\o_1^5\o_3^2+\o_1^5\o_2^2+\o_3^2\o_2^5)\Big]
+\mbox{permutations}\, .\nn\\
\eea
Note that the permutations in the above formula act on the ``objects''
$(\v1,\q1,\o_2)$, $(\v2,\q2,\o_3)$ and $(\v3,\q3,\o_1)$ (because of
the coupling of the proper times $\o_i$ and momenta $\q{i}$ in the
exponent of~(\ref{why})). 

Amongst these terms it is now instructive to focus on
a specific class of terms in the supergravity amplitude \eq{R4supergravity}.
We choose to study terms with the structure
\be
(\q{}\cdot\v{})^4\, \frac{v{}^4}{q^4}\, . 
\label{str}
\ee
On the Matrix Theory side these terms are easily isolated from 
$S^{3\rightarrow3}_Y|_{v^8}$ of \eq{why}, in particular
\be
S^{3\rightarrow3}_Y|_{v^4\, (\q{}\cdot\v{})^4}
=\Upsilon^2\,\int_0^\infty
\!d^3\o\,\frac{\Pi_2}{p^3}
\,\exp(-q_1^2 \o_2-q_2^2 \o_3 -q_3^2 \o_1)\, .
\ee
Of course it is rather difficult to perform this integral exactly. Being
interested only in the poles $1/(q_1^2\, q_2^2)$ and permutations thereof we 
proceed as follows. First perform the integral over (say) $\o_3$ exactly
and thereafter expand the integrand in powers of $1/\o_1$ and $1/\o_2$.
Using
\be
\int_0^\infty d\o \, \frac{1}{\o}\, e^{-\o\, q^2}= -\log q^2
-\gamma
\ee
we obtain the final result contributing to the structure \eqn{str}
(up to overall factors, dropping the
logarithms)
\be
S^{3\rightarrow3}_{2\rm\, loop}|_{v^4\, (\q{}\cdot\v{})^4}=
\Upsilon^2\, (\q{1}\cdot\v{12})^2\, \Big (\frac{1}{q_2^2\, q_3^2}
+\frac{1}{q_1^2\, q_2^2}\Big ) + \mbox{cyclic}
\ee
which is astonishingly close, but nevertheless {\it not} equal to 
the corresponding
terms in the supergravity amplitude of \eq{R4supergravity}
\be
{\cal A}_{{\cal R}^4}|_{v^4\, (\q{}\cdot\v{})^4}=
\Upsilon^2\, (\q{1}\cdot\v{12})^2\, \frac{1}{q_3^2\,q_1^2} + \mbox{cyclic}
\, .
\ee
This constitutes the abovementioned definite disagreement of the two results
and concludes our study of the ${\cal R}^4$ contributions to the
three graviton amplitudes.

\section{Conclusions}

In this work we have presented detailed comparisons between three
graviton scattering amplitudes in Matrix Theory and $d=11$
supergravity along with its leading M-theoretic curvature corrections.
On the one hand we have been able to complete and unify the results
of~\cite{oy,oy1} showing that the leading order $v^6$ eikonal spin 
independent $S$-matrices of tree level supergravity and two loop
Matrix Theory exactly agree. On the other hand, the moment one studies
the next to leading order $v^8$ Matrix Theory amplitude, the result
fails to match the corresponding (conjectured) term in ${\cal
R}^4$-corrected supergravity. Why does such a mismatch occur ?

In trying to answer this most pressing of questions, let us begin 
by noting that
our results pertain most strongly to the Susskind
finite $N$ formulation of the Matrix Theory conjecture~\cite{suss}.
Susskind's conjecture has be proven to be literally true in \cite{seib},
i.e. M-theory on a light-like circle with $N$ units of compactified
momentum 
is described by $U(N)$ Matrix Theory.
The real issue is what it implies for comparison with $d=11$
supergravity.
M-theory on a lightlike circle is Lorentz equivalent
to M-theory on a vanishing spacelike circle~\cite{seib}. 
On the contrary, supergravity
is a good effective description of M-theory at low energy and
at the same time when the radius of compactification is large
(so that all possible wrapped membranes are decoupled).
In terms of the string coupling constant $g_S$, for instance,
this shows that perturbative Matrix Theory and supergravity computations
are really trust-able in two different regions (respectively at small and large
values of $g_S$)\footnote{Roughly speaking this is due to the fact
that Matrix Theory is a good description of physics at substringy distances, 
whereas
supergravity is a good description at long wavelengths.}. 
It is then evident that no agreement should be expected {\it a priori}, except
for 
those amplitudes which are somehow protected from receiving 
any correction as one moves
from one regime to the other.
In view of the agreement found for tree level two and three particle scattering
amplitudes, this appears to be the case for the
terms of order $v^4$ and $v^6$ in the Matrix Theory effective
action as has been shown in \cite{pss} for the $U(2)$ and $U(3)$ models.
{}From this viewpoint the finite $N$ Matrix Theory conjecture, extended to the
supergravity regime, would require the existence 
of an infinite number of non-renormalization theorems.
However, our two loop order $v^8$ result indicates that there exists 
{\it no} non-renormalization theorem for these terms in the super Yang-Mills
quantum mechanics.

The underlying
type IIA string theory itself can be employed to understand the
relationship between perturbative Matrix Theory and low energy
$M$-theory. In particular the extensive agreement of one
loop spin dependent terms for $2\rightarrow 2$ scattering can
be easily understood by considering the scale independence of the
string theory cylinder/annulus amplitude between two D0 
particles \cite{scr,psw,psw1}. Indeed arguments
supported also by the string theory picture suggest that, 
if visible perturbatively, the
effects due to the ${\cal R}^4$ term may correspond to a  
five-loop non-planar contribution in Matrix 
Theory \footnote{Although again
the agreement would require a very non-trivial matching of amplitudes
computed in different regimes.}~\cite{ms}.
On the other hand
there is no perturbative ``string derivation'' of a 
correspondence between the next-to-leading $v^8$ two loop term
and the ${\cal R}^4$ amplitude given in (\ref{R4supergravity}).

Aside from the possibility of discovering new
non-renormalization theorems and although there 
were no real expectations for an agreement between two-loop 
Matrix theory and ${\cal R}^4$ supergravity corrections, neither was
there a definitive argument or computation to rule it out. 
We believe that our work gives a final (negative) answer to this question.
 
Finally, an obvious question to ask is whether one should find further
agreements with tree level supergravity.
Interestingly enough, in light of the simple Feynman diagrammatic
understanding of semi-classical recoil effects given in this work,
further comparison between Matrix Theory and tree-level
supergravity amplitudes can be contemplated for four graviton
scattering (i.e. three-loop level in the quantum mechanical model).
In \cite{dine2} it has been argued that at this order disagreement
is possible, but there is no definite answer yet.
As we have seen within our formalism of comparing directly S-matrices, 
it is quite easy to single out particular tensorial sub-structures.
In this way the analysis could
be greatly simplified and yet remain conclusive.

\vskip .8cm

\section*{Acknowledgments}

We thank Arkady Tseytlin for a most useful correspondence.
Moreover we wish to thank Wolfgang Junker, Sven Moch and
Washington Taylor for useful discussions.


\end{document}